# A Big Data Architecture for the Detection of Anomalies within Database Connection Logs

Swapneel Mehta, Prasanth Kothuri, Daniel Lanza Garcia

European Organisation for Nuclear Research

Meyrin, Geneva

{swapneel.sundeep.mehta, prasanth.kothuri, daniel.lanza}
@cern.ch

**Abstract.** We propose a big data architecture for analysing database connection logs across different instances of databases within an intranet comprising over 10,000 users and associated devices. Our system uses Flume agents sending notifications to Hadoop Distributed File System for long-term storage and Elasticsearch and Kibana for short-term visualisations, effectively creating a data lake for the extraction of log data. We adopt machine learning models with an ensemble of approaches to filter and process the indicators within the data, and aim to predict anomalies or outliers using feature vectors built from this log data.

## 1  Introduction

The project is to build a scalable and secure and central repository capable of storing consolidated audit data comprising listener, alert and OS log events generated by database instances. This platform will be used for extraction of data in order to filter outliers utilising machine learning approaches. The reports will provide a holistic view of activity across all oracle databases and the alerting mechanism will detect and alert on abnormal activity including network intrusion and usage patterns [1]. Database connection logs are analysed to flag potentially anomalous or malicious connections to the database instances within the network of the European Organisation for Nuclear Research (CERN). We utilise this research to shed light on patterns within the network in order to better understand the temporal dependencies and implement them in the decision-making process within the CERN system. These models are trained on subsets of a data lake that comprises daily connection logs across all instances of databases on the network. The data lake comprises Javascript Object Notation (JSON) logs that may be visualised using short-term storage, Elasticsearch, Grafana and Kibana or pushed to long-term storage on Hadoop Distributed File Storage (HDFS). We extract subsets of this data and apply models that vary based on different parameters of the data, and enable us to classify the outliers among the dataset as anomalies.

## 1.1 Database Instances in the CERN Network

**Fig 1.** Overview of the Database System at CERN [2]

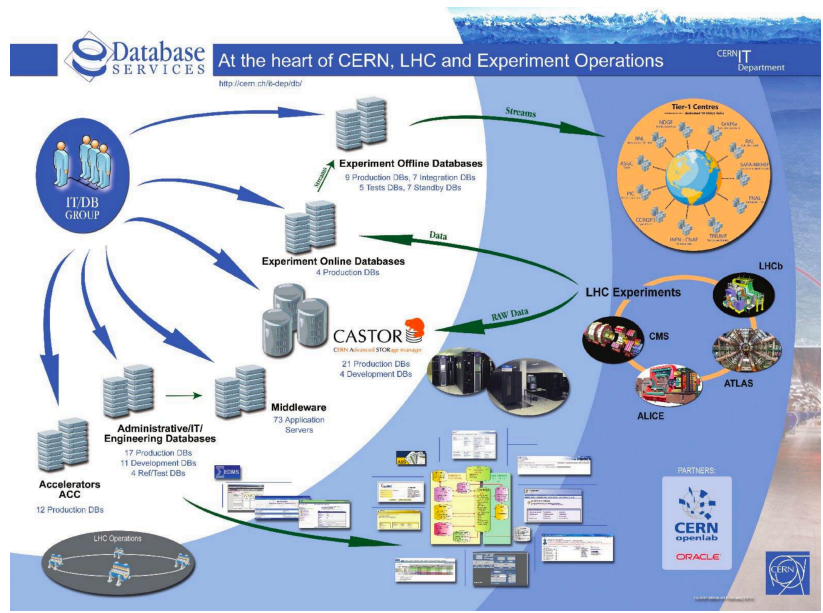

CERN follows an open access policy for most of its internal systems which extends to a significant percentage of database instances across the organisation. While this reduces the rule-based access control load on the system, it leads to a number of novel issues including but not limited to an unnecessarily large number of connection requests, connections accessing more resources than necessary, misused or abused credentials and more security-oriented issues such as malware causing repetitive connections from a host machine.

The CERN network comprises 10,000 active users and 1,500 different users [1] each with devices, including personal and officially supplied equipment. While most of the software installed on the systems is regulated and updates are managed centrally, there is considerable privilege afforded to a user to manage device personally and consequently the loopholes for the installation of user-defined software or scripts. These can end up being not only inefficient but also malicious in their operation, raising multiple issues within the system. Specifically, these systems might be compromised and utilised to either initiate multiple (spam) connections or use stolen credentials with malicious intent. There are numerous security issues with such a setup if there is unmonitored access to database instances. It can be used in multiple manners to throw the network into disarray due to the load on servers. We aim to use some concrete approaches in analysing these database connections by building models that can accurately fit this data and classify further connection requests by evaluating the likelihood of being anomalous. The architecture used to store this log data needs to have a low footprint and support real-time analytical pipelines reliant on time-series and sequence-based approaches.

## 2   Log Data Storage

**Fig 2.** Current Architecture for Storage of Log Data [3]

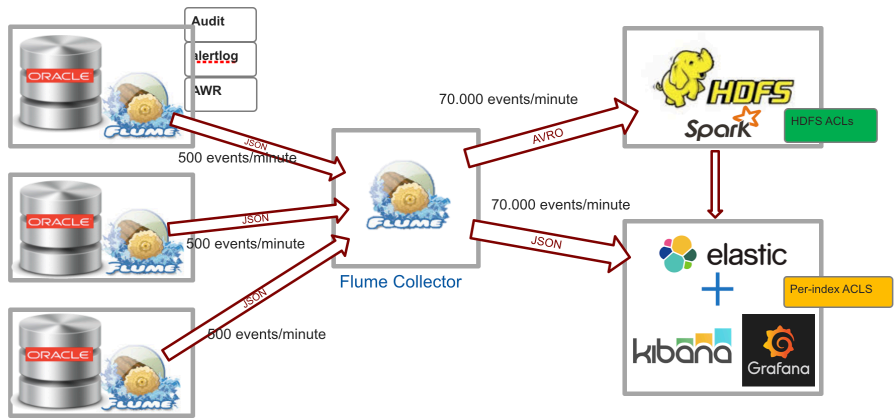

Database connection logs comprise multiple parameters per successfully established or failed connection which are utilised in building the feature vector for analysis. The architecture comprises of multiple instances of Oracle databases each of which have an Apache Flume agent running on it. Apache Flume is a reliable and distributed service for efficiently collecting, clustering, and transporting large amounts of data, often comprising system logs. The system architecture for this service encapsulates the concepts of data streaming flows. Flume offers reliability due to the arsenal of failover and recovery mechanisms incorporated into a tuneable set of options. The simplified data model allows for extensible online analytical implementation. It offers unprecedented scalability, reliability and performance [4]. Flume supports sources such as Avro, Thrift or logs, and a variety of sinks. It runs in the background, eliminating the need for scheduling jobs, and ensures all updates are streamed to HDFS as the configuration permits. In case of systems that do not load data directly, Flume provides an integration that allows for CSV and log data to be supported. [5]. Oracle Real Application Clusters (RAC) is an Oracle Database version involving clustering of instances across the system. It provides a comprehensive, high-availability stack that can be used as the foundation of a database cloud system as well as a shared infrastructure, ensuring high availability, scalability, and agility for any application [6]. The log data is transported via a central Flume Collector to Elasticsearch and Kibana for short-term storage primarily meant for data visualisation and long-term storage on HDFS.

## 2.1 Log Data Fields

The log consists of a number of different fields each with a value corresponding to the details of the connection. Some of the important connection details include fields such as timestamp, client_program (software that the client uses to connect), client_host (host from where the client connects; not necessarily specified), client_ip (IP address of the client), client_port (port used to connect), client_protocol (network protocol that the client uses to initiate connection), client_user (username provided to initiate connection), and so on.

**Fig 3.** Extracting User Connection Details to Build a Feature Vector

| hour_of_the_day | day_of_the_week | client_user | client_host | client_ip | client_program | CONNECT_DATA_INST | service_name |
|---|---|---|---|---|---|---|---|
| 11.821111111111112 | 1 | merge | pcamsj2.cern.ch | 137.138.188.167 | python | INT11R2 | int11r.cern.ch |
| 11.82611111111111 | 1 | | | | | | INT6R1 |
| 11.821111111111112 | 1 | merge | pcamsj2.cern.ch | 137.138.188.167 | python | INT11R2 | int11r.cern.ch |
| 11.826666666666666 | 1 | | | | | | INT11R2 |
| 11.821111111111112 | 1 | merge | pcamsj2.cern.ch | 137.138.188.167 | python | INT11R2 | int11r.cern.ch |
| 11.82611111111111 | 1 | merge | pcamsj2.cern.ch | 137.138.188.167 | python | | int11r.cern.ch |
| 11.821111111111112 | 1 | merge | pcamsj2.cern.ch | 137.138.188.167 | python | INT11R2 | int11r.cern.ch |
| 11.82611111111111 | 1 | | | | | | INT6R1 |
| 11.821111111111112 | 1 | merge | pcamsj2.cern.ch | 137.138.188.167 | python | INT11R2 | int11r.cern.ch |
| 11.826388888888889 | 1 | | | | | | INT11R1 |
| 11.821111111111112 | 1 | merge | pcamsj2.cern.ch | 137.138.188.167 | python | INT11R2 | int11r.cern.ch |
| 11.826388888888889 | 1 | | | | | | INT11R1 |
| 11.821111111111112 | 1 | merge | pcamsj2.cern.ch | 137.138.188.167 | python | INT11R2 | int11r.cern.ch |
| 11.821388888888889 | 1 | merge | pcamsj2.cern.ch | 137.138.188.167 | python | INT11R2 | int11r.cern.ch |

## 3 Anomaly Detection

**Fig 4.** An Overview of Strategies for Outlier Detection [7]

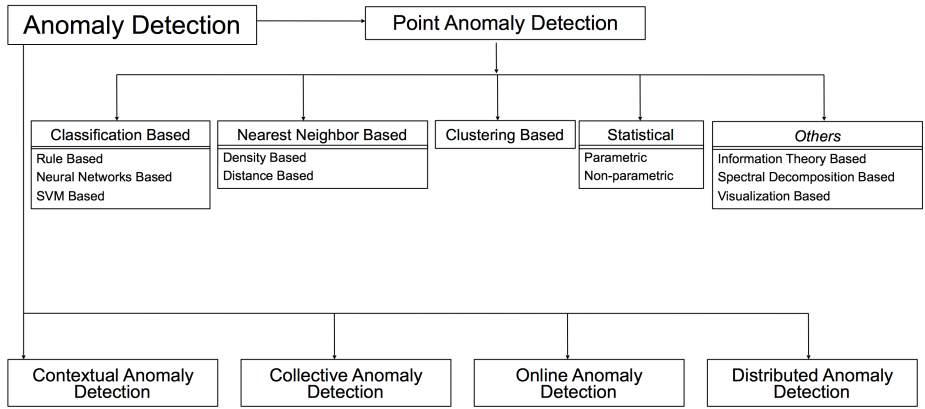

Anomaly Detection is the process of identifying unusual or outlying data from a given dataset. It has been a subject of interest as the amount of data available for analysis rises disproportionately as compared to the amount of available tagged data. Specifically, we utilise features extracted from established connection logs in order to determine the outliers for a given subset of temporal data obtained from a data lake. The data inherently possesses some degree of contamination which is why we set a clustering threshold as a cutoff for percentage of outliers within a data slice.

Models employed include K-Nearest neighbours, K-Means Clustering, Isolation Forests, Local Outlier Factor, and One-Class Support Vector Machines. Our results are based on a common subset of anomalies obtained from a majority of distance, density and classification approaches utilised for outlier detection.

**Table 1.** Comparison of Thresholds for Outliers across Models.

| Approach | Title | Outliers |
| --- | --- | --- |
| Distance | K-Nearest Neighbours | 2% |
| Distance | K-Means Clustering | - |
| Density | Isolation Forests | 3% |
| Density | Local Outlier Factor | 5% |
| Classification | One-Class SVM | 2% |

## 4  Conclusion

Hadoop and its assortment of components provide a well-established solution for data warehousing and provide a reliable pipeline for the transport, storage and streaming of log data produced by control systems at CERN. Spark is a system that has proven to be scalable in streaming applications when used in tandem with Hadoop. Further, it supports querying in SQL that simplifies the learning curve for users, should such a use-case arise. From our tests we find that the choice of the storage engine is very important for the overall performance of the system. The system can efficiently handle data synchronisation and perform near real-time change propagation.

**Acknowledgments.** The authors would like to acknowledge the contributions of Mr. Eric Grancher, Mr. Luca Canali, and other members of the CERN IT-DB Group towards this project.